\documentclass[a4paper,aps,pra,preprintnumbers]{revtex4}
\usepackage[utf8]{inputenc}
\usepackage[english]{babel}
\usepackage[T1]{fontenc}
\usepackage{amssymb,amsfonts,amsmath,mathtext,enumerate,float,dsfont}
\usepackage{graphics,graphicx,epsfig,epstopdf}
\usepackage{caption}
\usepackage{cmap}
\usepackage{multirow}
\usepackage{indentfirst}
\usepackage[usenames]{color}
\usepackage{amsthm}
\usepackage{xcolor}

\renewcommand{\Im}{\mathop{\mathrm{Im}}\nolimits}
\renewcommand{\Re}{\mathop{\mathrm{Re}}\nolimits}

\begin{document}

\title{Estimation of the set of states obtained in particle number measurement schemes}

\author{S. B. Korolev$^{1,2}$, E. N. Bashmakova$^{1}$, T. Yu. Golubeva$^{1,2}$} 
\affiliation{$^1$ St.Petersburg State University, Universitetskaya nab. 7/9, St.Petersburg, 199034, Russia}
\affiliation{$^2$ Laboratory of Quantum Engineering of Light, South Ural State University, pr. Lenina 76, Chelyabinsk, 454080, Russia}

\date{\today}

\begin{abstract}
The paper investigated a set of non-Gaussian states generated by measuring the number of particles in one of the modes of a two-mode entangled Gaussian state. It was demonstrated that all generated states depend on two types of parameters: some parameters are responsible for Gaussian characteristics, while other parameters are responsible for non-Gaussian characteristics. Among all generated states, we identified those optimally generated in terms of the generation probability and the magnitude of non-Gaussianity.
\end{abstract}

\maketitle

\section{Introduction}
There is currently a heightened interest in non-Gaussian states. This interest is driven by several factors. Firstly, the advantages of employing quantum non-Gaussian states as a resource have already been demonstrated in many areas, including quantum metrology \cite{Zhang, Hou,PhysRevResearch.3.033250}, quantum cryptography \cite{Lee2019} and information transmission \cite{ Guo2019}. Furthermore, non-Gaussian states allow improvements in existing quantum optics and informatics protocols. For instance, in the problem of quantum teleportation in continuous variables, by using auxiliary non-Gaussian states, errors can be significantly reduced \cite{PhysRevA.61.032302,Zinatullin2023,Zinatullin2021}. Conversely, the necessity for non-Gaussian states is also explained by the fact that to build universal quantum computing in continuous variables, one must be able to implement at least one non-Gaussian operation \cite{Braunstein_2005, Lloyd_1999}. A further significant factor contributing to the growing interest in non-Gaussian states in quantum computing is using such states in quantum error correction protocols \cite{Ralph_2003,Hastrup_2022}.

Among the numerous non-Gaussian states, a few have known applications. For instance, it is known that in various information applications (primarily in error correction protocols), Schr\"{o}dinger cat states \cite{Sychev2017,bashmakova2023effect,Ourjoumtsev2007,PhysRevA.87.062115}, GKP (Gottesman-Kitaev-Preskill) states \cite{Vasconcelos:10, PhysRevA.64.012310}, squeezed Fock states \cite{olivares2005squeezed,Kral1990, PhysRevA.109.052428}, and others are employed. However, to date, no general approaches and principles have defined the relationship between the characteristics of the non-Gaussian states themselves and their applications. Furthermore, theoretical protocols operate with idealized states, the generation of which, in most cases, is carried out only approximately \cite{Winnel,Podoshvedov_2023,Takase2021}.

The complete set of quantum non-Gaussian states cannot be classified using a single control parameter \cite{PRXQuantum.2.030204}. This presents a challenge that can be called the "classification problem of quantum non-Gaussian states". Currently, the classification of non-Gaussian states lacks a clear structure, in contrast to Gaussian states \cite{PRXQuantum.2.030204,PhysRevLett.123.043601}. Nevertheless, based on various measures and criteria for deviation from Gaussian statistics, which are usually called non-Gaussianity \cite{PhysRevA.90.013810}, it is possible to define different classes of non-Gaussian states. Non-Gaussianity can be quantified from various perspectives. Typically, Wigner negativity \cite{PhysRevA.87.062104} is employed for this purpose, in conjunction with measures based on the distance between reference Gaussian states and quantum non-Gaussian states (for example, the Hilbert–Schmidt distance \cite{PhysRevA.76.042327}, Hellinger distance \cite{Marian_2015}, Bures distance \cite{bures1969extension}, and others). It should be noted that there is no correlation between the measures describing non-Gaussianity and the field of application of non-Gaussian states. 

In the work, we focus on the classification of non-Gaussian states generated by a specific scheme. We evaluate the complete set of non-Gaussian states generated from a two-mode entangled Gaussian state by measuring the number of particles in one of the modes \cite{PhysRevA.109.052428}. Having established a relationship between the parameters of a two-mode entangled Gaussian state and the physical parameters of the output non-Gaussian state, we can identify two characteristic parameters corresponding to the generated states. One of the parameters is responsible for the structure of the generated non-Gaussian state itself, while the other is associated with Gaussian transformations. Furthermore, we demonstrate how the generation efficiency of non-Gaussian states can be quantitatively related to the magnitude of their non-Gaussianity.

The paper is structured as follows. Section II proposes a classification of the set of non-Gaussian states generated in a scheme for measuring a two-mode Gaussian entangled state by a particle number detector in terms of the parameters of the Gaussian transformations and the parameter responsible for the type of non-Gaussian state. Section III evaluates the efficiency of generating output states in a scheme from the point of view of the non-Gaussianity. Finally, Section IV determines the class of non-Gaussian states generated in the considered scheme with the highest probability.

\section{Output states}
The purpose of our work is to investigate the states obtained in schemes involving the measurement of particle numbers in one of the modes of a two-mode entangled Gaussian (TMEG) state. We want to comprehend how the physical parameters of these schemes impact the resulting state. Our objective is to identify the types of states that can be generated in such schemes, and to determine how to select the scheme parameters to obtain specific states. In essence, we seek to assess the complete set of states that can be generated by measuring the number of particles in one of the modes of a TMEG state.
 
As a general TMEG state, we consider a state whose wave function has the following form \cite{Rendell2005}:
\begin{align}
\label{TMGS}
    \Psi \left(x_1,x_2 \right)=\frac{\left(\mathrm{Re}[ a] \mathrm{Re} [c]-\left(\mathrm{Re }[b] \right)^2\right)^{1/4}}{\sqrt{\pi}}\exp \left[{-\frac{1}{2} \left(a x_1^2+2 b x_1 x_2+c x_2^2\right)}\right],
\end{align}
where  $a,b,c  \in \mathbb{C}$ satisfy the following conditions
\begin{align}
 \label{param_TMSS}
  \mathrm{ Re}[a] >0, \qquad   \mathrm{ Re}[c] >0, \qquad  \mathrm{Re} [a] \mathrm{Re}[ c]-\left(\mathrm{Re }[b] \right)^2 >0.
\end{align}
Physically, this state corresponds to the case when two arbitrarily squeezed vacuum states are entangled with each other. In this case, the type of entangling operation is not restricted in any way. This could be, for example, a beam splitter transformation \cite{Takase2021} or a CZ transformation \cite{PhysRevA.103.062407,PhysRevA.98.042304,PhysRevA.102.042608}. Hereafter, all wave functions are written in terms of the eigenvalues of the quadrature operator defined by the following relation $\hat{x}=(\hat{a}+\hat{a}^\dag)/\sqrt{2}$ where $\hat{a}$ stands for the annihilation operator of the quantum field. The subscript 1 and 2 specify the two modes.

We are interested in the state generated in mode 2 by the measurement of  number of particles in mode 1. In paper 1, we derived the exact analytical expression for the output state obtained by measuring the number of particles in one of the modes of a TMEG state, for arbitrary parameters $a$, $b$ and $c$. This state has the following form:
\begin{align} \label{Psi_out}
\Psi_n^{out}\left(x;a,b,c\right)=
\frac{\sqrt[4]{\mathrm{Re} [a] \mathrm{Re} [c]-\left(\mathrm{Re }[b] \right)^2}}{\sqrt[4]{\pi }\sqrt{ 2 ^n n! P_n} }
\sqrt{\frac{2(a-1)^n}{(a+1)^{n+1}}}
e^{-\frac{1}{2} x^2
\left(c-\frac{b^2}{a+1}\right)}
 H_n\left[\frac{b x}{\sqrt{a^2-1}}\right],
\end{align}
where $H_n$ stands for the Hermite polynomials \cite{Babusci2012}  defined such that the three  lowest of them read
\begin{equation}
\label{Her}
H_0(x) = 1, \qquad  H_1(x) = 2x,\qquad \mathrm{and}  \qquad H_2(x) = 4x^2-2,
\end{equation}
and $P_n$ is the probability of measuring the n-particles in mode 1, which is given by the following expression:
\begin{align}
 \label{norm}
    P_n=
   \frac{2| b| ^{2n} \sqrt{\mathrm{Re}[ a] \mathrm{Re} [c]-\left(\mathrm{Re}[ b] \right)^2}}{\left(|1+a| ^2 \mathrm{Re} \left[c-\frac{b^2}{a+1}\right]\right)^{n+1/2}} \,
   _2F_1\left[\frac{1-n}{2},-\frac{n}{2};1;\left| 1-\frac{\left(a^2-1\right) \mathrm{Re} \left[c-\frac{b^2}{a+1}\right]}{b^2}\right| ^2\right].
\end{align}
Here $_2F_1(x,y;s;t)$ is a hypergeometric function.

In order to evaluate the complete set of output states produced by the measurement of a TMEG state by a particle number detector, it will be useful for us to simplify the wave function of the output state. To do this, let us first rewrite the wave function using the new parameters:
\begin{align}
   \Psi_n^{\text{out}}\left(x;R,z\right)=\frac{\sqrt[4]{\Re(R)} | 1-z| ^{n/2} e^{-\frac{R x^2}{2}} H_n\left[x
   \sqrt{\frac{\Re[R]}{1-z}}\right]}{\sqrt[4]{\pi } \sqrt{2^n n!} \sqrt{ \,
   _2F_1\left(\frac{1-n}{2},-\frac{n}{2};1;| z| ^2\right)}},
\end{align}
where the new parameters are related to the parameters of TMEG states by the following relations:
\begin{align} \label{param_W_R}
  z=1-\frac{\left(a^2-1\right)}{b^2}\Re\left[c-\frac{b^2}{a+1}\right], \quad R=c-\frac{b^2}{a+1}.
\end{align}
This substitution may look quite strange now, but later, we will understand that the parameters introduced in this way have a transparent physical meaning. 

From the obtained expression, we can already conclude that the wave function depends not on three complex parameters $a$, $b$, and $c$, but on their combinations, which can be rewritten in the form of two complex parameters $z$ and $R$. This means that there always remains one free complex parameter (for definiteness, we assume it is parameter $a$). Since the output state does not depend on this parameter, by varying it, we can optimize the process of generating output states. The free parameter can be selected, for instance, to simplify the experimental scheme for generating output states. Or, as was shown in \cite{PhysRevA.109.052428,bashmakova2023effect}, by choosing the parameter $a$ we can maximize the probability of generating specific states. Thus, we can say that one of the parameters of the TMEG state affects only the magnitude of the probability of generating certain output states (determined by the parameters $R$ and $z$).

For further analysis of the output states, it will be useful for us to decompose the wave function of the output state into well-known states. In \cite{PhysRevA.109.052428}, it was demonstrated that accurate squeezed Fock (SF) states can be obtained at the output of the scheme if one chooses certain scheme parameters. This means it is most logical to decompose the output state as a superposition of SF states. Such a decomposition is given by the following equation:
\begin{align} \label{out_rw}
    \Psi_n^{\text{out}}\left(x;R,z\right)= \begin{cases}
    \hat{\mathcal{S}}\left(r_R e^{i\varphi_R}\right)\hat{\mathcal{R}}\left(-\frac{\arg \left[z\right]}{2}\right)\left(\sum \limits_{m=0}^{n/2}  A_{2m}\left(n,\left|z\right|\right) \Psi_{2m}^{\mathrm{F}}\left(x\right)\right), \quad \text{if} \quad n \text{ -- even},\\
    \\
     \hat{\mathcal{S}}\left(r_R e^{i\varphi_R}\right)\hat{\mathcal{R}}\left(-\frac{\arg \left[z\right]}{2}\right)\left(\sum \limits_{m=0}^{(n-1)/2}  A_{2m+1}\left(n,\left|z\right|\right) \Psi_{2m+1}^{\mathrm{F}}\left(x\right)\right), \quad \text{if} \quad n \text{ -- odd}.
    \end{cases}
\end{align}
where 
\begin{align}
\Psi_m^{\mathrm{\mathrm{F}}}(x)=\frac{e^{-\frac{1}{2} x^2} H_n(x)}{\sqrt{2^n n! \sqrt{\pi}}}   
\end{align}
is the wave function of the $m$-th Fock state \cite{Kral1990,NIETO1997135,PhysRevA.40.2494}, 
$\hat{\mathcal{S}}\left(re^{i\varphi}\right)=\exp\left[\frac{r}{2}\left(e^{i\varphi}\left(\hat{a }^\dag\right)^2-e^{-i\varphi}\left(\hat{a}\right)^2\right)\right]$ is a squeezing operator, and $\hat{\mathcal{R}} \left(\phi\right)=\exp\left[i\phi \hat{n}\right]$ is a rotation operator. The relation between the parameters of the squeezing operator and the parameter $R=\Re[R]+i\Im[R]$ of the output state is defined as follows:
\begin{align}
    &\cosh r_R=\frac{| R+1| }{2 \sqrt{\Re[R]}},\quad \tan \varphi_R=\frac{2 \Im[R]}{|R| ^2-1}.
\end{align}
The decomposition coefficients in Eq. (\ref{out_rw}) have the following form:
\begin{align}
    A_{m}\left(n,|z|\right)= \frac{\left| z\right| ^{\frac{n-m}{2}}}{ 2^{\frac{n-m}{2}} \left(\frac{n-m}{2}\right)! \sqrt{\,
   _2F_1\left(\frac{1-n}{2},-\frac{n}{2};1;\left| z\right| ^2\right)}}\sqrt{\frac{n!}{m!}}, 
\end{align}

From the Eq. (\ref{out_rw}) we can conclude that the output state is a finite (limited above by the number of measured particles $n$) superposition of Fock states with a certain parity (depending on the parity of the measured number of particles $n$). The decomposition coefficients depend only on the absolute value of $z$. By changing this value, we change the type of state generated. In addition, from Eq. (\ref{out_rw}), one can notice that the superposition of Fock states is acted by the rotation and the squeezing operators. The rotation operator depends only on the argument of $z$, and the squeezing operator depends only on the R parameter. For this reason, we can conclude that the R parameter is the parameter responsible for squeezing the output state. 

Thus, we have shown that the wave function of the output state is formed from two independent parts: a non-Gaussian superposition of Fock states and Gaussian transformations applied to this superposition. Moreover, the Gaussian transformation itself is a combination of rotation and squeezing operators.

\section{Non-Gaussianity of the output state}
\subsection{Non-Gaussianity parameter of the output state}
After we have assigned physical meaning to all parameters of the output state, we can proceed to evaluate the efficiency of generating certain states in the scheme under consideration. We can assess generation efficiency using two criteria. The first criterion is the non-Gaussian \cite{PhysRevA.90.013810} output state. Second criterion is the magnitude of the probability of state generation.

Let us begin our analysis with an estimation of the magnitude of the non-Gaussianity of the output states. The main physical essence of the generation scheme is that we project the mode of the TMEG state to a non-Gaussian Fock state (any non-vacuum Fock state). As a result, due to entanglement, we have a non-Gaussian state at the output of the scheme (except in the case of vacuum measurement). Roughly speaking, the non-Gaussianity obtained by the measurement in one mode is teleported to the other. As shown in \cite{PhysRevA.109.052428,bashmakova2023effect}, the efficiency of non-Gaussianity transfer in a similar scheme depends on the parameters of the scheme. In other words, for some generated states, the non-Gaussianity will be larger, while for others, it will be smaller. Our objective here is to identify the states with the greatest non-Gaussianity.

When discussing the non-Gaussianity of states, it is important to clarify what measure of non-Gaussianity is under consideration. There are numerous approaches for estimating the magnitude of non-Gaussianity \cite{PhysRevA.87.062104,PhysRevA.76.042327,Marian_2015,bures1969extension}. In the work, we will compare a few main ones and identify which output states exhibit the greatest non-Gaussianity according to various measures.

Before evaluating the non-Gaussianity of the output states, let us note that the magnitude of the non-Gaussianity of the state should be invariant under Gaussian transformations \cite{PRXQuantum.2.030204}. This means that the non-Gaussianity of our output states is equivalent to the non-Gaussianity of the superposition states:
\begin{align} \label{out_NG}
    \Psi_n^{\text{NG}}\left(x;\left|z\right|\right)= \begin{cases}
    \sum \limits_{m=0}^{n/2}  A_{2m}\left(n,\left|z\right|\right) \Psi_{2m}^{\mathrm{F}}\left(x\right), \quad \text{if} \quad n \text{ -- even},\\
    \\
     \sum \limits_{m=0}^{(n-1)/2}  A_{2m+1}\left(n,\left|z\right|\right) \Psi_{2m+1}^{\mathrm{F}}\left(x\right), \quad \text{if} \quad n \text{ -- odd}.
    \end{cases}
\end{align}
As noted earlier, this state depends only on the absolute value of the parameter $z$. This means the non-Gaussian state depends only on the parameter $|z|$. In what follows, we will call the parameter $|z|$ the non-Gaussianity parameter of the output state.

\subsection{Wigner negativity of the output state}

One commonly used measure of non-Gaussianity is the Wigner negativity (WN). This measure shows the magnitude of the negative part of the Wigner function. The larger the negative part is, the greater the non-Gaussianity of the state under consideration. It is important to note that this measure has a significant drawback. If we use it to study a non-Gaussian state with a positive Wigner function, this measure will not distinguish it from a Gaussian state. However, in our case, this disadvantage is not critical. The fact is that all output non-Gaussian states obtained in our scheme have the negative part of the Wigner function. This is because all wave functions are proportional to the Hermite polynomial, which has a negative part. Therefore, in our case, the WN can be considered a measure of the non-Gaussianity of the output states.

\begin{figure}[H]
    \centering
    \includegraphics[scale=0.9]{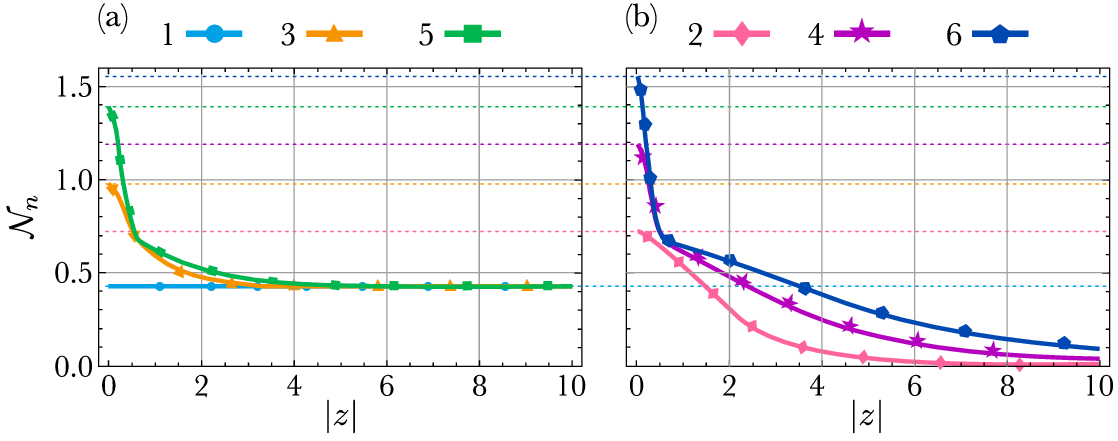}
    \caption{Dependence of the WN value of the output state on the non-Gaussianity parameter $|z|$. On the graph, different colors indicate curves corresponding to different numbers of measured particles $n$. Case (a) corresponds to measuring an odd number of particles, and case (b) corresponds to an even number of particles. The dotted lines indicate the maximum WN for different cases of measuring $n$.}
    \label{fig:Neg}
\end{figure}

The WN of the output state is defined by the following expression \cite{Kenfack_2004}:
\begin{align}
    \mathcal{N}_n=\int \int  \left|W_n^{\text{out}}\left(x,p\right)\right|dx dp-1,
\end{align}
where $W_n^{\text{out}}\left(x,p\right)$ is the Wigner function of the output state. Fig. \ref{fig:Neg} shows the dependence of the WN of the non-Gaussian part of the output state (\ref{out_NG}) depending on the non-Gaussianity parameter $|z|$.
\noindent
Several conclusions can be drawn from the graph. First, the maximum WN is achieved for all $n$ when the parameter $|z|=0$. In this case, the non-Gaussianity of the output state increases monotonically with the number of measured particles. Substituting the value $|z|=0$ into Eq. (\ref{out_NG}) allows us to conclude  that the state with the highest magnitude of non-Gaussianity will be the Fock state:
\begin{align} 
\Psi_n^{\text{NG}}\left(x;0\right)=\Psi_n^{\mathrm{F}}\left(x\right).
\end{align}

Furthermore,  from Fig. \ref{fig:Neg} (b), it is clear that at large $|z|$, the WN of states with an even number $n$ tends to zero. The WN of states with an odd number of measured particles tends to $0.426$ (see Fig. \ref{fig:Neg} (a)). Passing to the limit $|z|\rightarrow \infty $ in Eq. (\ref{out_NG}), we find that all states with even numbers tend to the vacuum state, while states with odd numbers tend to the first Fock state:
\begin{align} 
\Psi_n^{\text{NG}}\left(x;|z|\rightarrow \infty\right)=\begin{cases}
    \Psi_0^{\mathrm{F}}\left(x\right), \quad \text{if}\quad n \;\text{-- even}, \\
    \Psi_1^{\mathrm{F}}\left(x\right), \quad \text{if}\quad n \;\text{-- odd}.
\end{cases}
\end{align}
Moreover, from Fig. \ref{fig:Neg} (a), one can notice that the magnitude of the non-Gaussianity of the state with $n=1$ remains constant. This is because when measuring one particle in the mode of a TMEG state, we always obtain the first squeezed Fock state in the other mode \cite{olivares2005squeezed,PhysRevA.109.052428}.

Thus, the analysis of the magnitude of WN allows us to conclude that the maximum non-Gaussianity will be achieved for the generation of the Fock state. As Gaussian transformations were not considered in Eq. (\ref{out_NG}), it can be concluded that among all output states, squeezed rotated Fock states are generated most efficiently in terms of non-Gaussianity.

However, this conclusion is limited by the fact that the value of WN is determined through the integral of the modulus of the Wigner function. Such an integral can only be calculated numerically for arbitrary parameters $|z|$ and $n$. Consequently, it is only possible to provide numerical estimates for a limited number of parameters. It is necessary to consider an alternative measure to identify states with the highest magnitude of non-Gaussianity analytically.

\subsection{Measure of non-Gaussianity based on the distance between Gaussian and non-Gaussian states}
As such a measure, let us consider a measure based on the distance in Hilbert space between the non-Gaussian state of interest and some Gaussian state. As a Gaussian state, we use a state whose first and second moments coincide with those of the non-Gaussian state of interest. The greater the distance between two states, the more the non-Gaussian state differs from the Gaussian one, and thus the higher the magnitude of the non-Gaussianity. To evaluate the non-Gaussianity of our state, we will employ a measure based on the fidelity of the Gaussian and non-Gaussian states as a distance measure \cite{Ghiu_2013}. This measure is given by the following expression: 
\begin{align} \label{dist_NG}
    \mathcal{M}\left(\hat{\rho}_n^{\text{NG}},\hat{\rho}_G\right)=1-\sqrt{\text{Fid}\left(\hat{\rho}_n^{\text{NG}},\hat{\rho}_G\right)},
\end{align}
where $\hat{\rho}_n^{\text{NG}}$ is the density matrix of our non-Gaussian state (\ref{out_NG}), and $\hat{\rho}_G$ is the density matrix of the Gaussian states with the first and second moments equal to the moments of the non-Gaussian state under interest.

The explicit expressions for the moments of the output state, as well as for the fidelity of the output non-Gaussian state and the Gaussian state with the same moments, are presented in Appendix \ref{app_MNBF}. Here, let us plot the dependence of the magnitude of non-Gaussianity (\ref{dist_NG}) on the parameter $|z|$. Fig. 2 shows the graph of this dependence. 
\begin{figure}[H]
    \centering
    \includegraphics[scale=0.9]{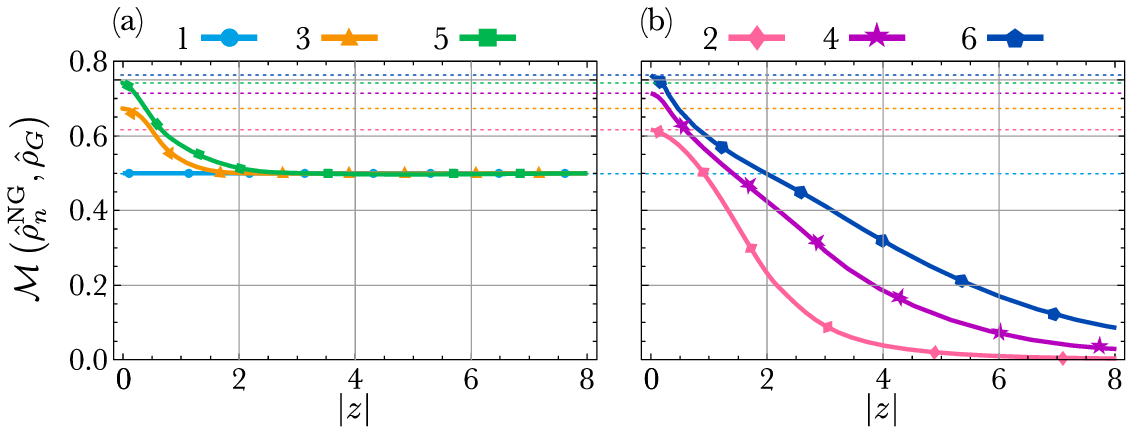}
    \caption{Dependence of the value of the measure of non-Gaussianity of the output state $ \mathcal{M}\left(\hat{\rho}_n^{\text{NG}},\hat{\rho}_G\right)$ on the non-Gaussianity parameter $|z|$. On the graph, different colors indicate curves corresponding to different numbers of measured particles $n$. Case (a) corresponds to measuring an odd number of particles, and case (b) corresponds to an even number of particles. The dotted lines indicate the maximum values of $\mathcal{M}\left(\hat{\rho}_n^{\text{NG}},\hat{\rho}_G\right)$ for different cases of measuring $n$.}
    \label{fig:enter-label}
\end{figure}
\noindent
The figure illustrates that the magnitude of the non-Gaussianity of the output state depends on the non-Gaussianity parameter $|z|$. It can be seen that this behaves similarly to the magnitude of the WN. The maximum non-Gaussianity is reached at $|z|=0$, while the minimum is attained at $|z| \to \infty$. As for WN, the non-Gaussianity computed using the measure (\ref{dist_NG}) decreases monotonically from the maximum to the minimum. Using the distance-based non-Gaussian measure, we can prove analytically (see Appendix \ref{app_MMNG}) that the maximum of the non-Gaussianity of the output state will be at $|z|=0$, i.e., when the squeezed Fock state is generated.

The coincidence of the results for the two considered measures, as well as the fact that for pure states all other known non-Gaussianity measures are associated with the distance measure \cite{zhang2020quantifying}, means that the result we obtained does not depend on the choice of the non-Gaussianity measure. In the scheme under consideration, the squeezed Fock state is generated most efficiently in terms of non-Gaussianity. This result has a very simple physical explanation. As mentioned earlier, in our scheme, we teleport the non-Gaussianity obtained from the Fock state measure to the unmeasured mode. This means that the maximum level of non-Gaussianity in the output state is limited by the non-Gaussianity of the measured Fock state. It follows that we cannot exceed the non-Gaussianity introduced into the scheme, and it is the non-Gaussianity of the Fock state. Moreover, the Gaussian transformations do not affect the magnitude of the non-Gaussianity. This means that the scheme most efficiently generates squeezed rotated Fock states.

\section{Most probable output states}
Let us now evaluate the efficiency of generating output states in terms of their probability. As it is known \cite{Takase2021,PhysRevA.109.052428,Podoshvedov_2023}, in schemes involving the measurement of particle numbers, different output states are generated with different probabilities. The subject of this section is to identify, among the set of output states (\ref{out_rw}), those that are generated with the highest probability. To do this, let us rewrite the probability of measuring a certain number of particles (\ref{norm}) using the parameters we have introduced (\ref{param_W_R}):
\begin{multline} \label{norm_AZ}
    P_n\left(z,a\right)=\sqrt{\frac{
   4 \Re\left[a\right] | 1-z| -2| a-1|  \left(| a+1| -\Im\left[a-1\right] \sin \theta-\Re\left[a-1\right] \cos \theta \right)}{| a+1|^2  | 1-z| }}*\\
   \left| \frac{a-1}{(a+1) (1-z)}\right| ^n \, _2F_1\left(\frac{1-n}{2},-\frac{n}{2};1;| z| ^2\right),
\end{multline}
where, for simplicity, the notation is introduced:
\begin{align}
    \theta=\arg \left[a-1\right]-\arg \left[1-z\right].
\end{align}

We see that the probability function depends on two parameters. The first of these is parameter $z$, while the second is parameter $a$, which, as previously noted, is solely responsible for the probability of generating states. In this case, the probability does not depend on the parameter $R$, i.e., it does not depend on the squeezing degree of the output state. Consequently, for each state characterized by parameter $z$, it is possible to optimize the probability by selecting parameter $a$. This probability is independent of whether the state is squeezed or not. 

Let us now find the states for which the generation probability is maximal. To this end, we may consider the maximum of the function defined in Eq. (\ref{norm_AZ}). When measuring an odd number of particles, the global probability maximum occurs when two conditions are met: $z=0$ and the real and imaginary parts of the parameter $a$ lie on circles of the form: 
\begin{align} \label{rings}
    \left(a''\right)^2+\left(a'-\left(n+\frac{1}{4 n+2}+\frac{1}{2}\right)\right)^2=\frac{4 n^2 (n+1)^2}{(2 n+1)^2},
\end{align}
where $a'$ and $a''$ are the real and imaginary parts of the parameter $a=a'+ia''$. For the case of measuring an even number of particles $n$, the probability will have global and local maxima. The global maximum of the probability is reached at $z\to \infty$. However, as we have already found out, this case corresponds to vacuum state generation. The local maximum of the probability is reached at $z=0$ and when the real and imaginary parts of the parameter $a$ satisfy the Eq. (\ref{rings}). This maximum coincides with the maximum for the case of measuring an odd number of particles.

The solutions obtained indicate that the maximum probability will be achieved when generating Fock states. Moreover, considering that the probability is invariant for the squeezing parameter $R$, squeezed Fock states also have the highest probability.

\section{Conclusion} 
In this work, we addressed the problem of generating different quantum states by measuring the number of particles in the mode of a two-mode entangled Gaussian state. A two-mode entangled Gaussian state is characterized by three complex parameters: $a$, $b$ and $c$. We could specify a change of variables that allowed us to move from the parameters of a two-mode entangled Gaussian state to the physical parameters of the output state $\lbrace a,b,c \rbrace \rightarrow \lbrace a,z,R \rbrace$. We demonstrated that the type of output state depends only on two parameters: $z$ and $R$. The free parameter $a$ can be controlled to optimize the output state generation scheme. These results allow one to construct an output state with specified properties directly.

We demonstrated that the output state can be represented as a superposition of a finite number of Fock states, which are acted upon by two Gaussian transformations: a rotation transformation and a squeezing transformation. The parameter $R$ is solely responsible for the squeezing transformation of the output state, while $\arg [z]$ is responsible for the rotation of the superposition state. The superposition depends only on the absolute value $|z|$. In other words, the parameter $|z|$ is responsible for the generated non-Gaussian state. Therefore, for the scheme under consideration, we can classify the non-Gaussianity of the output states based on one parameter that naturally arises in the analytical description and is directly related to the input characteristics of the scheme.

We associated the non-Gaussianity parameter we introduced with known non-Gaussianity measures. Based on this, we could estimate the effectiveness of generating output states in terms of the non-Gaussianity magnitude. To assess non-Gaussianity, we considered two measures: Wigner negativity and a measure based on the distance between the state under study and a Gaussian state with the same first and second moments. Analyzing these two measures, we showed that, among all the output states obtained by measuring a certain number of particles $n$, the squeezed Fock states will have the highest non-Gaussianity.

Finally, we investigated the efficiency of generating states in terms of their probability. We demonstrated that the squeezed Fock states not only maximize non-Gaussianity but are also generated with the highest probability. 

Thus, we can conclude from the above that the scheme generates squeezed Fock states most efficiently. Any other states generated in the considered scheme have lower non-Gaussianity and lower generation probability. In particular, the presented scheme generates squeezed Schr\"{o}dinger cat states less efficiently.

Since the considered scheme represents a rare case where a whole class of non-Gaussian states is generated accurately, it can be directly used in error correction protocols based on squeezed Fock states \cite{QECCSFOCK}. Moreover, it can also be used in protocols where squeezed Fock states are used to generate more exotic non-Gaussian states \cite{Winnel}.

\section*{Funding}
This research was supported by the Theoretical Physics and Mathematics Advancement Foundation "BASIS" (Grants No. 24-1-3-14-1 and No. 24-1-5-162-1). SBK and TYG acknowledge support by the Ministry of Science and Higher Education of the Russian Federation on the basis of the FSAEIHE SUSU (NRU) (Agreement No. 075-15- 2022-1116).

\section*{Disclosures}
The authors declare no conflicts of interest.

\section*{Data availability} Data underlying the results presented in this paper are not publicly available at this time, but may be obtained from the authors upon reasonable request.

\bibliography{nongaussian}

\appendix
\section{Fidelity-based measure of non-Gaussianity} \label{app_MNBF}
The fidelity-based measure of the non-Gaussianity is defined as follows:
\begin{align} 
    \mathcal{M}\left(\hat{\rho}_n^{\text{NG}},\hat{\rho}_G\right)=1-\sqrt{\text{Fid}\left(\hat{\rho}_n^{\text{out}},\hat{\rho}_G\right)},
\end{align}
where $\hat{\rho}_n^{\text{NG}}$ is the density matrix of our non-Gaussian state (\ref{out_NG}), and $\hat{\rho}_G$ is the density matrix of the Gaussian states with the first and second moments equal to the moments of the non-Gaussian state under interest.

To estimate the magnitude of non-Gaussianity of the output state, it is necessary to calculate its first and second moments. As the state is non-displaced, its first moments are zero. The second moments of the non-Gaussian state are given by the following expressions:
\begin{align}
    &d_x= \langle \delta x^2 \rangle=\frac{1+2n}{2}+\frac{n (n-1) |z|(1-|z|) \, _2F_1\left(\frac{3-n}{2},1-\frac{n}{2};2;|z|^2\right)}{2 \,
   _2F_1\left(\frac{1-n}{2},-\frac{n}{2};1;|z|^2\right)},\\
       &d_p= \langle \delta p^2 \rangle=\frac{1+2n}{2}-\frac{n(n-1)|z| (|z|+1) \,
   _2F_1\left(\frac{3-n}{2},1-\frac{n}{2};2;|z|^2\right)}{2 \,
   _2F_1\left(\frac{1-n}{2},-\frac{n}{2};1;|z|^2\right)}.
\end{align}
Given these expressions, we can write the fidelity of our non-Gaussian state and the Gaussian state with the same first and second momenta in the form:
\begin{align} \label{m_fid}
 \text{Fid}\left(\hat{\rho}_n^{\text{out}},\hat{\rho}_G\right)=\sum _{m_1=n-2\left\lfloor\frac{n}{2}\right\rfloor}^{n} \sum _{m_2=n-2\left\lfloor\frac{n}{2}\right\rfloor}^{n} \frac{n!\left(\frac{z}{2}\right)^{n-\frac{m_1+m_2}{2}} \delta _{n-m_1,2 \left\lfloor
   \frac{n-m_1}{2}\right\rfloor } \delta _{n-m_2,2 \left\lfloor \frac{n-m_2}{2}\right\rfloor }}{\sqrt{m_1!}
   \sqrt{m_2!} \left(\frac{n-m_1}{2}\right)! \left(\frac{n-m_2}{2}\right)! \, _2F_1\left(\frac{1-n}{2},-\frac{n}{2};1;z^2\right)}F\left(m_1,m_2,d_x,d_p\right).
\end{align}
Here $\delta _{n-m,2 \left\lfloor \frac{n-m}{2}\right\rfloor }$ is the Kronecker delta, which is non-zero only if the parity is $n$ and $m$ is the same. In the Eq. (\ref{m_fid}), an auxiliary function of the form was introduced:
\begin{multline}
    F\left(n,m,d_x,d_p\right)=\frac{2\sqrt{m!} \sqrt{n!} \left(\frac{d_x-d_p}{(2d_x+1) (2d_p+1)}\right)^{\frac{m+n}{2}}
   \delta _{m-2 \left\lfloor \frac{m}{2}\right\rfloor ,n-2 \left\lfloor \frac{n}{2}\right\rfloor }}{\sqrt{(2d_x+1)
   (2d_p+1)}} *\\
   \sum _{k=0}^{\min (n,m)} \frac{(-1)^k}{k!} \left(\frac{(2d_x+1) (2d_p+1)}{d_x-d_p}\right)^k\;\sum _{q=0}^{\min (n,m)-k}\frac{1}{{q!}} \left(\frac{2d_x (2d_p+1)}{d_x-d_p}\right)^q\chi\left(n,m,k,q,d_x,d_p\right),
\end{multline}
where
\begin{align}
\chi\left(n,m,k,q,d_x,d_p\right)=\sum_{q_1=0}^{\min (n,m)-k-q}\frac{ \left(\frac{(2d_x+1) 2d_p}{d_x-d_p}\right)^{q_1} \delta _{m-k-q-q_1,2
   \left\lfloor \frac{m-k-q-q_1}{2}\right\rfloor } \delta _{n-k-q-q_1,2 \left\lfloor \frac{
   n-k-q-q_1}{2}\right\rfloor }}{q_1! \left(\frac{m-k-q-q_1}{2}\right)! \left(\frac{n-k-q-q_1}{2}\right)!}.
\end{align}

\section{Maximizing a measure of non-Gaussianity based on fidelity} \label{app_MMNG}
Let us select parameters to maximize the fidelity-based measure of non-Gaussianity:
\begin{align} 
    \mathcal{M}\left(\hat{\rho}_n^{\text{NG}},\hat{\rho}_G\right)=1-\sqrt{\text{Fid}\left(\hat{\rho}_n^{\text{NG}},\hat{\rho}_G\right)},
\end{align}
To maximize this function, we need to require that $\text{Fid}\left(\hat{\rho}_n^{\text{out}},\hat{\rho}_G\right)$ takes its smallest value. That is, we need to minimize the function (\ref{m_fid}). Since the difference
\begin{align}
    d_x-d_p=\frac{2 (n-1) n |z| \, _2F_1\left(\frac{3-n}{2},1-\frac{n}{2};2;|z|^2\right)}{\,
   _2F_1\left(\frac{1-n}{2},-\frac{n}{2};1;|z|^2\right)}\geqslant 0 
\end{align}
non-negative for any $|z|$ and $n$, then this means that the function $ F\left(n,m,d_x,d_p\right)$ is always non-negative. This means that sum 
\begin{align} 
 \text{Fid}\left(\hat{\rho}_n^{\text{out}},\hat{\rho}_G\right)=\sum _{m_1=n-2\left\lfloor\frac{n}{2}\right\rfloor}^{n} \sum _{m_2=n-2\left\lfloor\frac{n}{2}\right\rfloor}^{n} \frac{n!\left(\frac{z}{2}\right)^{n-\frac{m_1+m_2}{2}} \delta _{n-m_1,2 \left\lfloor
   \frac{n-m_1}{2}\right\rfloor } \delta _{n-m_2,2 \left\lfloor \frac{n-m_2}{2}\right\rfloor }}{\sqrt{m_1!}
   \sqrt{m_2!} \left(\frac{n-m_1}{2}\right)! \left(\frac{n-m_2}{2}\right)! \, _2F_1\left(\frac{1-n}{2},-\frac{n}{2};1;z^2\right)}F\left(m_1,m_2,d_x,d_p\right).
\end{align}
is the sum of non-negative terms. The minimum of this sum is reached when $|z|=0$. In this case, only one terms with $m_1=m_2=n$ remains from the whole sum. Thus, the minimum fidelity will be given by the expression: 
\begin{align}
\text{Fid}\left(\hat{\rho}_n^{\text{out}},\hat{\rho}_G\right)=F\left(n,n,d_x,d_p\right)=\frac{n^n}{(n+1)^{n+1}},
\end{align}
where we have taken into account that at $|z|=0$ the variances on $x$ and $p$ quadratures are equal to each other. That is 
\begin{align}
    d_x=d_p=\frac{2n+1}{2}.
\end{align}
Thus, the maximum of the fidelity-based measure of non-Gaussianity is given by:
\begin{align} 
    \max _{\hat{\rho}_n^{\text{NG}}} \left[\mathcal{M}\left(\hat{\rho}_n^{\text{NG}},\hat{\rho}_G\right)\right]=1-\sqrt{\frac{n^n}{(n+1)^{n+1}}}.
\end{align}
\end{document}